\documentstyle[10pt,aaspp4]{article}

\lefthead{von Braun et al.}
\righthead{Color-color Relations for Cluster Giants}

\begin{document}

\title{Color-color Relations for Red Giants in Star Clusters}

\author{Kaspar von Braun, Kristin Chiboucas, Jocelyn Kelly Minske, Jos\'{e} Francisco Salgado}

\affil{University of Michigan}
\authoraddr{Department of Astronomy, Univ. of Michigan, Ann Arbor, MI 48109-1090}
\authoremail{kaspar@astro.lsa.umich.edu, kristin@astro.lsa.umich.edu,
kelly@astro.lsa.umich.edu, salgado@astro.lsa.umich.edu}

\and
\author{Guy Worthey}
\affil{St. Ambrose University}
\authoraddr{Department of Physics and Astronomy, St. Ambrose University, 
518 W. Locust St., Davenport, IA 52803}
\authoremail{gworthey@saunix.sau.edu}

\begin{abstract}

New Johnson-Cousins $UBVRI$ photometry of giants in globular clusters
is combined with $JHK$ photometry on the CIT system 
to produce color sequences for
giants from the globular clusters M3, M5, M13, and M92. $UBVRI$
data are also presented for giants in the metal-rich open cluster NGC 6791.
These data fill a gap in the literature,
especially for the $R$ \& $I$ bands.  We provide the empirical 
relations between broad band colors for various [Fe/H] values for metal-poor
giants. 
The color sequences for $U-B$ and $B-V$ show clear separations 
for different [Fe/H] values.
We also find weak, though unexpected, metallicity dependences of $V-R$,
$V-I$, and $J-K$ colors. $H-K$ is metal-insensitive. The above colors
are plotted as a function of $V-K$, and a literature
$(V-K) - T_{\rm eff}$ relation is given. 

\end{abstract}


\section{Introduction}

Population synthesis models and stellar evolutionary isochrones 
are the tools which give age estimates for galaxies and star clusters,
thus providing a constraint on the age of the universe.
The models, however, 
are dependent upon the transformations from 
effective temperature ($T_{\rm eff}$), surface gravity,
and abundance 
to observed colors and magnitudes. 
In the case of metal-poor giants,
these transformations remain inconclusive, despite
considerable efforts in the past. 

In 1966,
Johnson\footnote{Johnson, H. L. 1966, \araa, 4, 193}\markcite{j66} 
produced a remarkable set of tables that
gives color-color sequences in $UBVRIJKLMN$ passbands for local
dwarfs, giants, and supergiants 
(of nearly solar metallicity, since metal-poor stars are rare in the disk),
based on his and
his collaborators' extensive
photometry efforts. He calculated the absolute flux in each 
passband to obtain bolometric corrections and effective temperatures. 
For giants, this temperature scale has been subsequently revised
(Dyck et al. 1996; DiBenedetto 1993; 
Ridgway et al. 1980)\markcite{dbvr96,d93,rjww80} 
but the color-color sequences
are still used in
many contexts, including assigning colors to theoretical isochrones
and computing colors for integrated light population models.  Since
1966, new passbands have become standard for broadband photometry.
These are the Cousins (1980a,b)\markcite{c80a,c80b} $R$ and $I$
filters at roughly 0.68$\mu$m and 0.79$\mu$m, and the CIT (Elias et
al. 1982; ``CIT'' stands for California Institute 
of Technology)\markcite{efmn82} $H$ filter at 1.6$\mu$m.  
With these new filters, it is clear that more comprehensive 
color-color tables are necessary. 

One powerful way to generate a complete color-color table is to
compute theoretical line-blanketed stellar spectra and convolve them with
filter response functions to synthesize colors (e.g. Buser \& Kurucz
1979; Bell \& Gustafsson 1978, 1989; Kurucz
1992).\markcite{bk1979,bg78,bg89,k92} The only major detriment to
this approach is that, owing to the complexity of the problems of
opacity and convection, the theoretical fluxes match imperfectly with
real stellar fluxes. Systematic color drifts can be seen when theoretical
results are compared to an
empirical color-color table (cf. Worthey 1994).\markcite{w94}

Empirical tables are therefore of primary interest, for use by
themselves or for use as a check of the theoretical color
calibrations. The most complete empirical calibration attempted to
date was done by Green (1988)\markcite{g88} for use in the Revised Yale
Isochrones (Green et al. 1987).\markcite{RYI}
For Green's work, solar
neighborhood $UBVRI$ photometry was assembled, and a temperature
scale was attached via the $R-I$ color. 
The extension to different
metallicities was accomplished by using theoretical color results.
The Green calibration is imperfect; while problems with
fitting isochrones (Corbally 1996)\markcite{c96} may be 
due to the temperature of
the evolutionary tracks rather than the color calibration, 
the Green colors differ from other literature
calibrations (Worthey 1994).\markcite{w94}
A problem with bolometric corrections (Tiede, Frogel, \& Terndrup
1995; Mould 1992)\markcite{tft95,m92} has also been noticed,
but the color calibration has not yet been redone.

As part of an effort to redo this color calibration, a literature
search conducted by Worthey and Fisher (1996)\markcite{wf96} 
discovered a paucity of observations in the $R$ and
$I$ passbands for metal-poor giants.
New observations at the Michigan-Dartmouth-MIT (MDM)
Observatory were proposed to help fill this $RI$ gap and 
to provide empirical relations between the various
colors for different metallicities.
We selected globular 
clusters of known metallicity, with previous $BV$ 
(and sometimes $U$) photometry,
and specifically stars observed in
$JHK$ by Frogel et al. (1983)\markcite{frogel83} 
and Cohen et al. (1978)\markcite{cohen78}. 
In order to extend the metallicity range of our data to 
higher [Fe/H] values, NGC 6791 was added to the list. 
Unreduced $JHK$ photometry exists and, we expect,
will soon be available to complete the 
color sequence for this cluster. 
The approach in this paper is to concentrate
{\it only} on the Frogel et al. (1983) and 
Cohen et al. (1978)\markcite{cohen78} stars in the globular clusters and on
NGC 6791 giants previously studied by Garnavich et
al. (1994).\markcite{gvzh94} In this way, we reduce a daunting array
of data to a manageable size while retaining full ability to construct
$UBVRIJHK$ color sequences from the results.

We describe the observations and data reduction in the next section.
In section 3.1, we give our photometry results and, 
in section 3.2, compare them 
with data found in the literature. Section 3.3 contains a discussion 
of the empirical color-color relations for the metal-poor giants. 
Section 4 contains concluding remarks.

\section{Observations and Data Reduction}

The observations were obtained during the nights of April 11, 12, and
13, 1997, at the MDM Observatory McGraw-Hill 1.3m
telescope. A Schott glass $UBVRI$ filter set fabricated to match the
Johnson-Cousins system was used with a
UV-coated Tek 1024 CCD. Landolt (1992)\markcite{land92} standard
stars were regularly observed along with fields that were
chosen to overlap previous Frogel et al. (1983)\markcite{frogel83}
and Cohen et al. (1978)\markcite{cohen78} $JHK$ target stars in
globular clusters, and Garnavich et al. (1994)\markcite{gvzh94} red
giants in open cluster NGC 6791.

The initial processing of the raw CCD images was done with the routines
in the IRAF\footnote{IRAF is distributed by the
National Optical Astronomy Observatories, which are operated by the
Association of Universities for Research in Astronomy, Inc., under
cooperative agreement with the NSF.} CCDPROC package. For each
of the three nights, 10 biases were combined for the bias subtraction.
The flats were produced by combining between 3 and 6
twilight flat images per night per filter.
 
The processed data were reduced using DoPHOT 
(Schechter et al. 1993).\markcite{schech93}
Because the point spread function (PSF) of the stars varied with position
on the CCD, we adjusted the values of some of the input parameters of 
DoPHOT to obtain accurate photometry results. In particular, we 
used the variable PSF feature of DoPHOT and, in some cases, 
lowered DoPHOT's sensitivity with respect to detecting nearby neighbors
of stars. The latter was done in order to avoid false multiple detections
of individual, slightly elongated stellar images.
 
DoPHOT's photometry output contains a list of aperture corrections for
stars with small photometric errors and without nearby neighbors.  In
order to apply the correct aperture correction for each of our target
stars, we selected from this list the stars far away from the cluster,
so as to avoid aperture corrections influenced by the higher sky value
in the immediate vicinity of the cluster. With the fairly large field
of view of the CCD, this method left us with between roughly 10 and
240 isolated stars with high S/N ratios for the various
images. Using these stars, the dependence (if any) of the aperture
correction upon position on the CCD was determined, using a linear fit
in $x$ and $y$.  The F-test (cf. Press et al. 1992)\markcite{press92}
was used to determine if fits which
included $x$- and $y$-terms were statistically different from a
constant value for the aperture correction applied over the whole
chip.
 
During our three photometric nights, we observed a total of 10 Landolt
(1992)\markcite{land92} fields, at airmass values ranging from $\sim
1.15$ to $\sim 1.55$.  Using the IRAF PHOTCAL package, we applied
standard star solutions for all three nights of the form

\begin{equation}
 V = v + a_{0} + b_{0} X_{v} + c_{0} (b-v)
\end{equation}
\begin{equation}
 B - V = a_{1} + b_{1} X_{b} + c_{1} (b-v)
\end{equation}
\begin{equation}
 U - B = a_{2} + b_{2} X_{u} + c_{2} (u-b)
\end{equation}
\begin{equation}
 V - R = a_{3} + b_{3} X_{r} + c_{3} (v-r)
\end{equation}
\begin{equation}
 V - I = a_{4} + b_{4} X_{i} + c_{4} (v-i),
\end{equation}

\noindent
where the $a_{j}$, $b_{j}$, and $c_{j}$ are the fitted constants, 
$X_{filter}$ is the airmass of the exposure taken with the respective
filter, the lowercase magnitudes are instrumental, and the uppercase ones
are the final calibrated magnitudes. 
 
\begin{figure}
\plotone{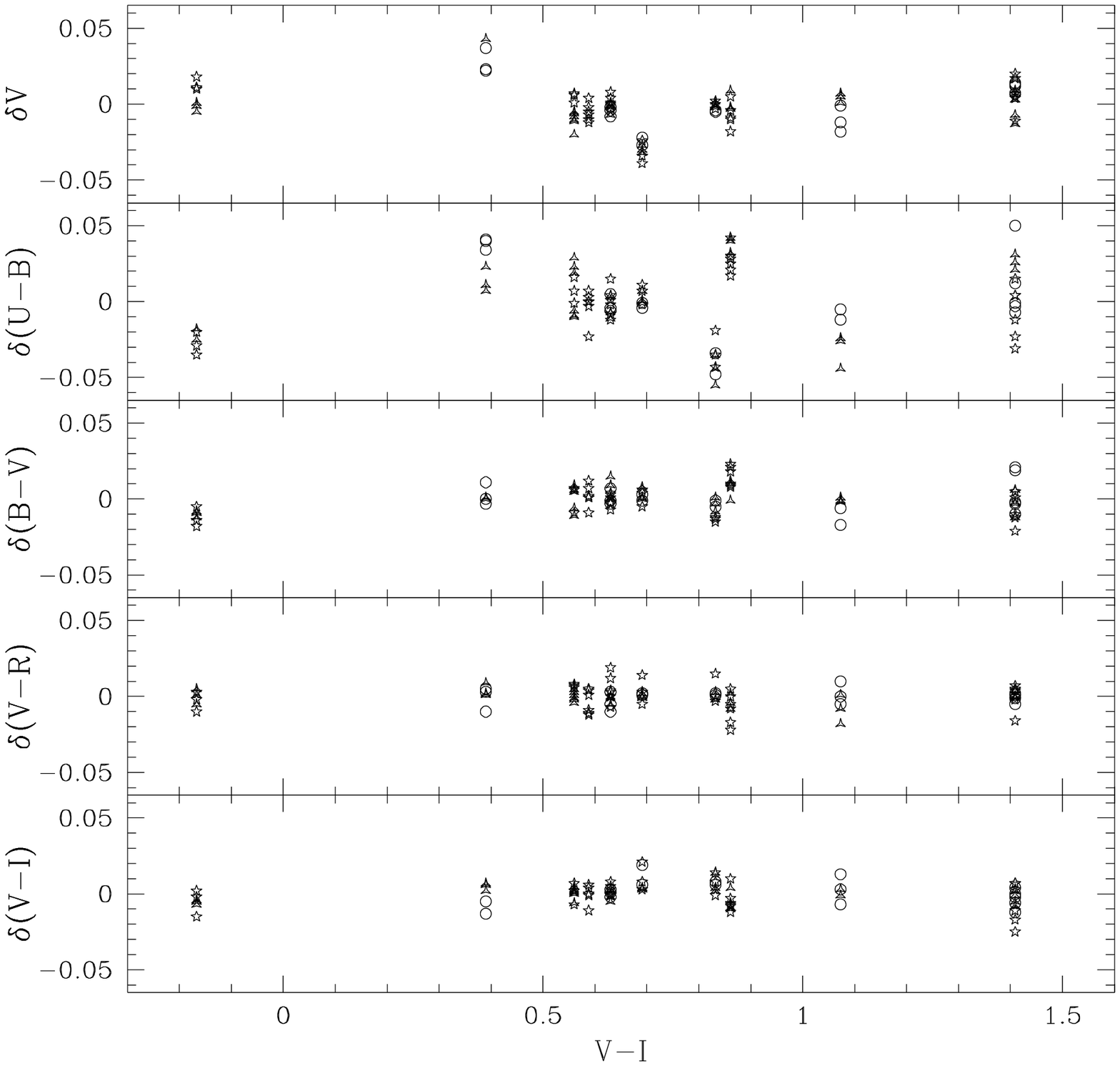}
\caption{The residuals between the calculated
magnitudes and standard magnitudes of the $\sim90$ Landolt (1992)
stars vs their standard (V-I) color. Open circles, triangles, and
stars respresent datapoints obtained during night 1, night 2, and
night 3, respectively.  \label{resids}}
\end{figure}

The RMS errors for the fits for the three nights are given in 
Table 1, and the residuals between the calculated magnitudes
and the ones given by Landolt (1992)\markcite{land92} 
are plotted in Fig.~\ref{resids}.
 
\placetable{table1} 
 
\placefigure{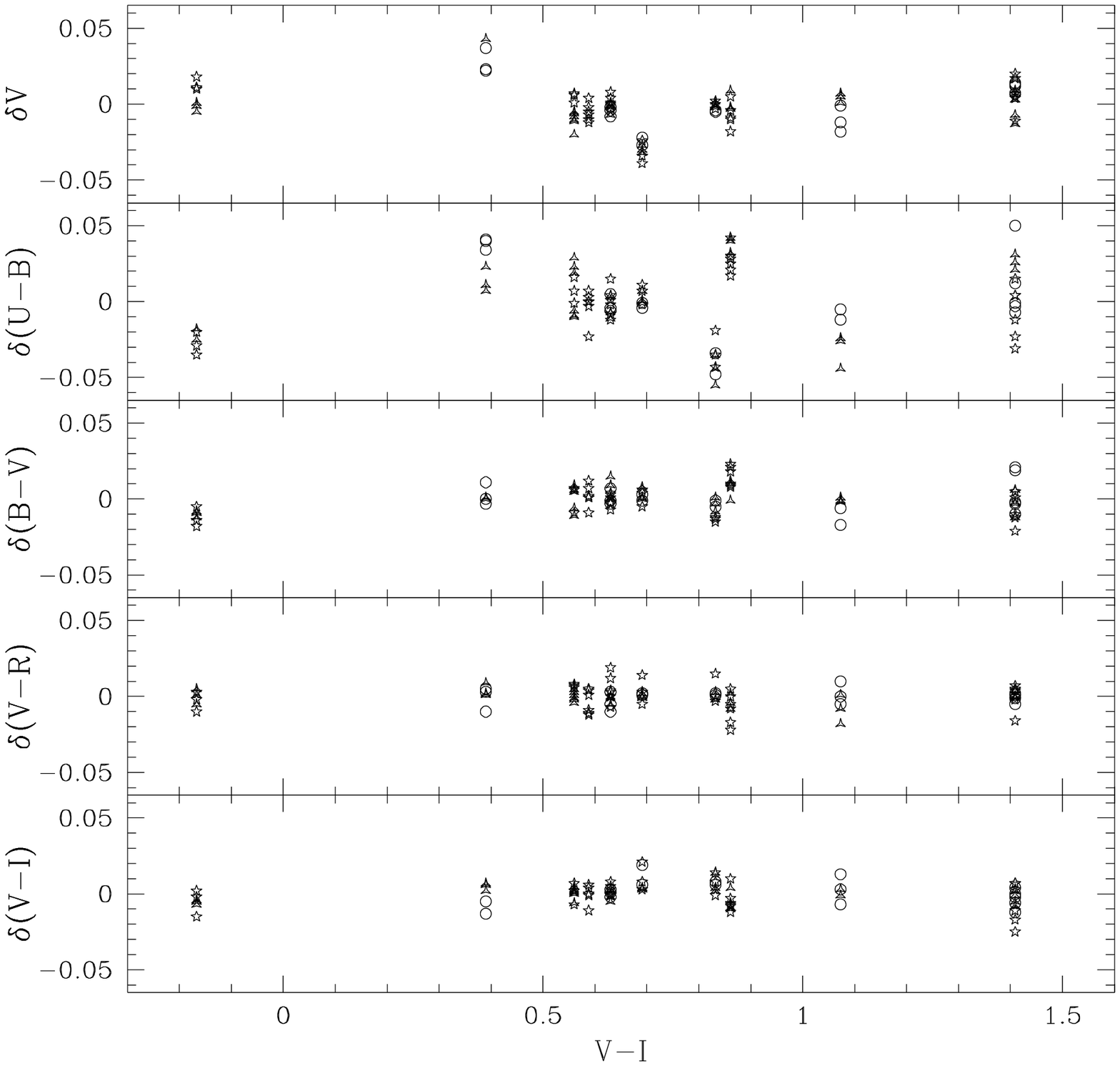}

Roughly two-thirds of the program stars were observed only on one
night (usually 3 CCD frames per filter, 2 frames for the $U$ filter), but
about one-third of the stars were observed on two nights.
For all the stars, the $V$ magnitudes and colors were averaged arithmetically.

The photometric scatter between nights 2 and 3, based on 8 stars
in NGC 6791, is around 0.01 mag for $VRI$ and about 0.015 mag
for $UBV$ colors.
With only 8 stars,
the estimated error of our photometry for these nights
is the photometric scatter divided by $\sqrt{8}$,
which is around 0.004 mag and 0.005 mag for $VRI$ and $UBV$ colors,
respectively. These values are consistent with the
RMS of the standard star solutions (Table 1).

However, for 7 globular cluster stars in common
between nights 1 and 2, the colors matched to a 
photometric scatter of 0.03 mag for all
colors, significantly larger than what the standard star solutions
imply. The $B-V$ colors for these stars
were also systematically different by 0.05 mag. For
these reasons, night 1 data were dropped whenever possible.  Only two
stars were observed in night 1 alone, and these are marked in
Table 2.

\section{Results and Discussion}

\subsection{Photometry Results}

\placetable{table2}

\placetable{table3}

Reduced magnitudes and errors for 
our program giants (luminosity class 3) are given in Table 2. 
The ``Err1'' column gives a magnitude error estimate based
on the number of observations and the scatter between individual 
measurements. The error is quantized in 0.005-mag steps. Colors involving
the $U$ filter should be assumed to have 50\% larger errors than the
``Err1'' entry would suggest. The ``Err2'' column
contains an entry if one or more filters had few observations or if
there was extra scatter between measurements. For instance, an entry
of ``.03ui'' means that any color involving $U$ or $I$ should be
considered to have an error of 0.03 mag.

Table 3
contains the dereddened (via Cardelli et al. 1989)\markcite{card89}
photometry results. Star identifications are
given along with [Fe/H] and $E(B-V)$ from Harris (1996)\markcite{h96}. 
Infrared photometry is from Cohen et al. (1978)\markcite{cohen78} 
and Frogel et al. (1983).\markcite{frogel83} Their
raw photometry results were used, also dereddened using Cardelli
et al. (1989).\markcite{card89}
This allows full $UBVRIJHK$ color sequences to be constructed.

\subsection{Literature Comparison}

We conducted a literature search to compare our photometry to 
previous studies whenever possible. 
For $U-B$, we found 20 globular cluster stars in common
with pre-1975 photometry (no more recent data were discovered). The
median $\delta (U-B)=(U-B)_{\rm this \ work} - (U-B)_{\rm literature} $
is $-0.08$. By contrast, three NGC 6791 stars in common with Harris \&
Canterna (1981)\markcite{hc81} compare with $\delta
(U-B)=+0.14$. It thus seemed that $\delta (U-B)$ increased with 
increasing [Fe/H] values,
and we considered various filter defects that could generate this metallicity
trend. However, the behavior is counter to that expected
from a red leak, and the UV coating on the CCD should make the overall
$U$ filter response fairly similar to the original, so we
believe that we have approximated the Landolt standard system very well,
as implied by the standard star solutions.

The other colors did not display such a trend. 
Literature $B-V$ values were found for 42 program stars
(Cathey (1974);\markcite{cathey1974}
Arp (1955);\markcite{arp55} 
Arp (1962);\markcite{arp62}
Kaluzny (1993, private communication);
Cudworth (1995, private communication);
Harris \& Canterna (1981);\markcite{hc81} 
Kinman (1965);\markcite{kinman65}
Kaluzny \& Rucinski (1995)\markcite{kal95}), 
and the overall average $\delta (B-V)
= -0.03$. Concentrating on the most recent data, the three
NGC 6791 stars in common with Harris \& Canterna (1981)\markcite{hc81} 
give $\delta (B-V)=+0.007$, and the three stars in
common with Kaluzny \& Rucinski (1995)\markcite{kal95} give $\delta
(B-V)=+0.015$. 
$VI$ photometry of 13 NGC 6791 stars in common with Garnavich et
al. (1994)\markcite{gvzh94} yields an average $\delta (V-I)=-0.15$ mag, but we
are less concerned with this offset because of the relatively low
accuracy of 0.05 mag claimed by these authors.

\subsection{Discussion}

The major goal of this study is to explicitly reveal the metallicity
dependence of broad-band colors. For instance, $(U-B)_{0}$ has long been
used to estimate the metallicity of faint stars (e.g. van den Bergh
1962).\markcite{vdb62} Figures \ref{ubvk}, \ref{bvvk}, \ref{vrvk},
\ref{vivk}, \ref{jkvk}, and \ref{hkvk} show the various dereddened
colors versus
$(V-K)_{0}$ for
giants of widely differing metal abundances, including 
the globular cluster giants from this work.
$(V-K)_{0}$ is chosen because it is an excellent
temperature indicator for GKM giants, with a large color range
compared to its observational uncertainty. In addition, the best model
fluxes show no metallicity sensitivity. That is, one temperature - $(V-K)_{0}$
conversion is applicable to stars of all abundances, at least in the
4000K to 5000K range (e.g. Bessell et al. 1989, 1991; Kurucz 1992; Bell \&
Gustafsson 1989).\markcite{bbsw89,bbsw91,k92,bg89}

A fit to the effective temperature --- $(V-K)_{0}$ relation of Ridgway
et al. (1980)\markcite{rjww80} is
\begin{equation}
(V-K)_0 = 34.19 - 0.01520 T + 2.420\times 10^{-6} T^2 
- 1.330\times 10^{-10} T^3 ,
\end{equation}
where $T$ is $T_{\rm eff}$ in degrees Kelvin. The range is valid over
$1.5 < (V-K)_0 < 4$, or $3800 < T_{\rm eff} < 5600$ K. This
approximate formula is given as a convenience for readers, but they
should be aware that the cubic curve can deviate from the Ridgway et
al. calibration by as much as about 0.08 mag in $(V-K)_0$, which
corresponds to about 60 K in effective temperature. Serious users may
want to refer to the original temperature calibration table.

Our photometry plus the Worthey \& Fisher (1996)\markcite{wf96}
literature photometry, all dereddened, produced relations summarized
here as six formulae giving colors as a function of $(V-K)_{0}$ and
[Fe/H]. Much of the literature photometry comes from the machine
readable version of Morel \& Magnenat (1978)\markcite{mm78} and is
on the Johnson system. The $RI$ data were 
transformed using the relations given by
M. S. Bessell (1979).\markcite{b79} 
[Fe/H] values were obtained from McWilliam (1990)\markcite{mcwilliam90}
and Cayrel de Strobel et al. (1992).\markcite{cds92}
The fits below are non-linear
least-squares regressions as described by 
Press et al. (1992).\markcite{press92} 
Some data were rejected in a 2.5-$\sigma$
rejection loop, but never more than a few percent were discarded.
Most of the literature data are photographic with quoted error of
$\sim 5 \%$, whereas the data of this work have errors more on
the order of $\sim 2 \%$. Consequently, we included a weighting factor
of $\sim 2.8$ to our data. This value is derived from the 
conservative estimate that our errors are up to 60\% of the 
literature errors.
Theoretical colors from Worthey (1994) were included at low
statistical weight ($\frac{1}{4}$ of the literature datapoints)
to provide guidance in regions with no stars. These
48 ``stars'' are included in the total N (number of datapoints)
below, except for the $(V-I)_{0}$ equation 
(theory data produce a large range in V-I for a given V-K which
is not reflected in observational data).  Different
combinations of coefficients were compared using the F-test and only
statistically significant terms were retained. The fits are good {\it
only} in the regime $ 1.6 < (V-K)_{0} < 3.6 $, but cover the 
range of Galactic [Fe/H] values for giants.
Note that, although we omit stars with $-1 < {\rm [Fe/H]} < 0$
in the plots for the sake of clarity, all metallicities are 
included in the derivation of the fits. 

\begin{eqnarray} 
(U-B)_{0} & = & \ -1.0121 \ + \ 0.8239 \ (V-K)_{0} \ + \ 0.04431 \ 
{\rm [Fe/H]}^{2} \nonumber\\
 & & + \ 0.3752 \ (V-K)_{0} \ {\rm [Fe/H]} \ - \ 0.0868 \ (V-K)_{0}^{2} \ {\rm
 [Fe/H]}; \\
  & & (N = 428, RMS = 0.082) \nonumber\\
 & & \nonumber\\
 (B-V)_{0}  &  = &  \ 0.1116 \ + \ 0.4013 \ (V-K)_{0} \ + \ 0.3509 \ 
{\rm [Fe/H]} \nonumber\\ 
   & & - \ 0.0823 \ (V-K)_{0} \ {\rm [Fe/H]} 
     + \ 0.01298 \ (V-K)_{0} \ {\rm [Fe/H]}^{2};\\
   & & (N = 625,  RMS = 0.062) \nonumber\\
 & & \nonumber\\
 (V-R)_{0}  & = &  \ -0.0040 \ + \ 0.2271 \ (V-K)_{0} \ + \ 0.0100 \ 
{\rm [Fe/H]};\\
 & & (N = 301, RMS = 0.021) \nonumber\\
 & & \nonumber\\
 (V-I)_{0}  & = & \ 0.1595 \ + \ 0.3387 \ (V-K)_{0} \ + \ 
0.004291 \ (V-K)_{0}^{3} \nonumber\\
 & & + 0.004255 \ (V-K)_{0}^{2} \ {\rm [Fe/H]} \ - \ 0.005596 \ {\rm [Fe/H]^{3}};\\
 & & (N = 267, RMS = 0.026) \nonumber\\
 & & \nonumber\\
 (J-K)_{0}  & = & \ 0.0231 \ + \ 0.2613 \ (V-K)_{0} \  + \ 0.009869 \ (V-K)_{0} 
\ {\rm [Fe/H]}; \\  
 & & (N = 642, RMS = 0.029)\nonumber\\
 & & \nonumber\\
 (H-K)_{0}  & = & \ -0.0190 \ + \ 0.04402 \ (V-K)_{0}; \\
 & & (N = 446, RMS = 0.020)\nonumber\end{eqnarray}
 
The $(V-R)_{0}$ fit contains
an [Fe/H] term that is statistically significant, but only
at the $50\%$ level. The
real behavior of $(V-R)_{0}$ is probably more subtle than the simple fit
can reproduce. 
The $(J-K)_{0}$ color equation contains a very
significant $(V-K)_{0}{\rm [Fe/H]}$ term. 
This indicates a metallicity dependence of the $(J-K)_{0}$ color, 
which is usually
assumed to be a pure temperature indicator. 
Fitting the $(V-I)_{0}$ data produced a somewhat surprising 
result. At first glance, one might assume that the data may 
be well represented by a metallicity-independent, linear fit in 
$(V-K)_{0}$ (see Fig.~\ref{vivk}). 
We find, however, a more complex behavior with 
respect to both metallicity and $(V-K)_{0}$.
The functional dependence of $(V-I)_{0}$ upon [Fe/H] is weak, 
but significant. 

There is a possibility that an error in the assumed $E(B-V)$ 
values of the various clusters could affect the fitting 
results. We investigated the effect an $E(B-V)$ error
of $\pm$0.02 would have on datapoints but found that the resulting shifts
in color are small compared to the spread due to different metallicities.
In order to demonstrate the magnitude and direction
of the shift of the datapoints in the color-color diagrams
due to a reddening error, we provide the reddening vectors
for $E(B-V)=0.10$ (see Figures~\ref{ubvk} through~\ref{hkvk}).
This value is, of course, far beyond a reasonable error in $E(B-V)$,
but lower values produced vectors which were too small 
to be visible in the figures.
The vectors were created using the reddening curve by 
Cardelli et al. (1989).\markcite{card89}

Our photometry is good enough to show separation of 
$(U-B)_{0}$ sequences for
globular clusters of differing metallicities in Figure~\ref{ubvk}. M92
stars (at [Fe/H]$=-2.29$)\footnote{All globular cluster [Fe/H] values 
are from Harris (1996).} lie to 
numerically smaller (i.e., bluer) $(U-B)_{0}$ from the other globular
clusters. There is also a hint that M3 ([Fe/H]$=-1.57$) 
and M13 stars ([Fe/H]$=-1.54$)
lie blueward of M5 ([Fe/H]$=-1.29$). This illustrates the metallicity
sensitivity of $(U-B)_{0}$ and also confirms our optimistic assessment of
our photometric accuracy. Our M92 data are significantly bluer
in $(U-B)_{0}$ than literature data of similar [Fe/H] in the range
$2.1 < (V-K)_{0} < 2.7$. Stars in this region have higher $T_{\rm eff}$ 
and are thus fainter since they lie further toward the subgiant
branch. It is likely that the discrepancy is due to the 
fact that the literature data are photographic, but 
more CCD data are needed to verify this.
Included in the figure are the 
$(U-B)_{0}$ fits for several [Fe/H] values. 

Figure~\ref{bvvk} also clearly displays a dependence of $(B-V)_{0}$ 
color upon [Fe/H].
It is interesting to note that all the fits for the various values
of metallicity intersect in the region $(V-K)_{0} \sim 3$. 
Figures~\ref{vrvk} and~\ref{jkvk} show a much weaker dependence of 
$(V-R)_{0}$ and $(J-K)_{0}$ 
on [Fe/H], as indicated by the equations above.  
There is a hint that the $(V-R)_{0}$ sequence shows a
divergence for different metallicity ranges
around $2 < (V-K)_{0} < 3$, but more data are needed to make
sure. 

The interesting behavior of $(V-I)_{0}$ is displayed in Figure~\ref{vivk}.
Giants with solar [Fe/H] fall toward numerically lower $(V-I)_{0}$
values than the ones with [Fe/H] $<2$ for $(V-K)_{0} < 2.4$. At  
$(V-K)_{0} > 2.8$, however, the solar metallicity curve lies above the 
ones representing the
lower metallicity ranges. Furthermore, stars with [Fe/H] $=-1$ 
fall toward lower $(V-I)_{0}$ than solar [Fe/H] stars throughout 
the range of $(V-K)_{0}$. Another noteworthy result is the non-linear
dependence of $(V-I)_{0}$ on $(V-K)_{0}$.
It is apparent from the 
plot that the fits approximate the datapoints very well. 
Since this $(V-I)_{0}$ behavior was unexpected, we examined the 
possibility of single datasets overly influencing the fit,
but when the function was refit without the Walker (1994)\markcite{walk94} 
and the da Costa \& Armandroff (1990)\markcite{da90} M15
data, as well as the theoretical points, it remained unchanged. 
It is worth mentioning that the dependence of $(V-I)_{0}$ 
on metallicity is a 
significantly lower-amplitude effect than the one on
$(V-K)_{0}$. $(V-I)_{0}$ therefore remains a useful color index for 
temperature determination. 

Figure~\ref{hkvk} shows the opposite case: $(H-K)_{0}$
displays no sensitivity to metal abundance. In this plot, stars of all
abundances overlie the same locus. It is most likely that both
colors, $(H-K)_{0}$ and $(V-K)_{0}$, have negligible metallicity
sensitivity (rather than the alternative that both have measureable
sensitivities, but the sensitivities conspire to look similar in the
color-color plot). The $(H-K)_{0}$ plot has a very small range
compared to its observational error.

\section{Concluding Remarks}

\begin{figure}
\plotone{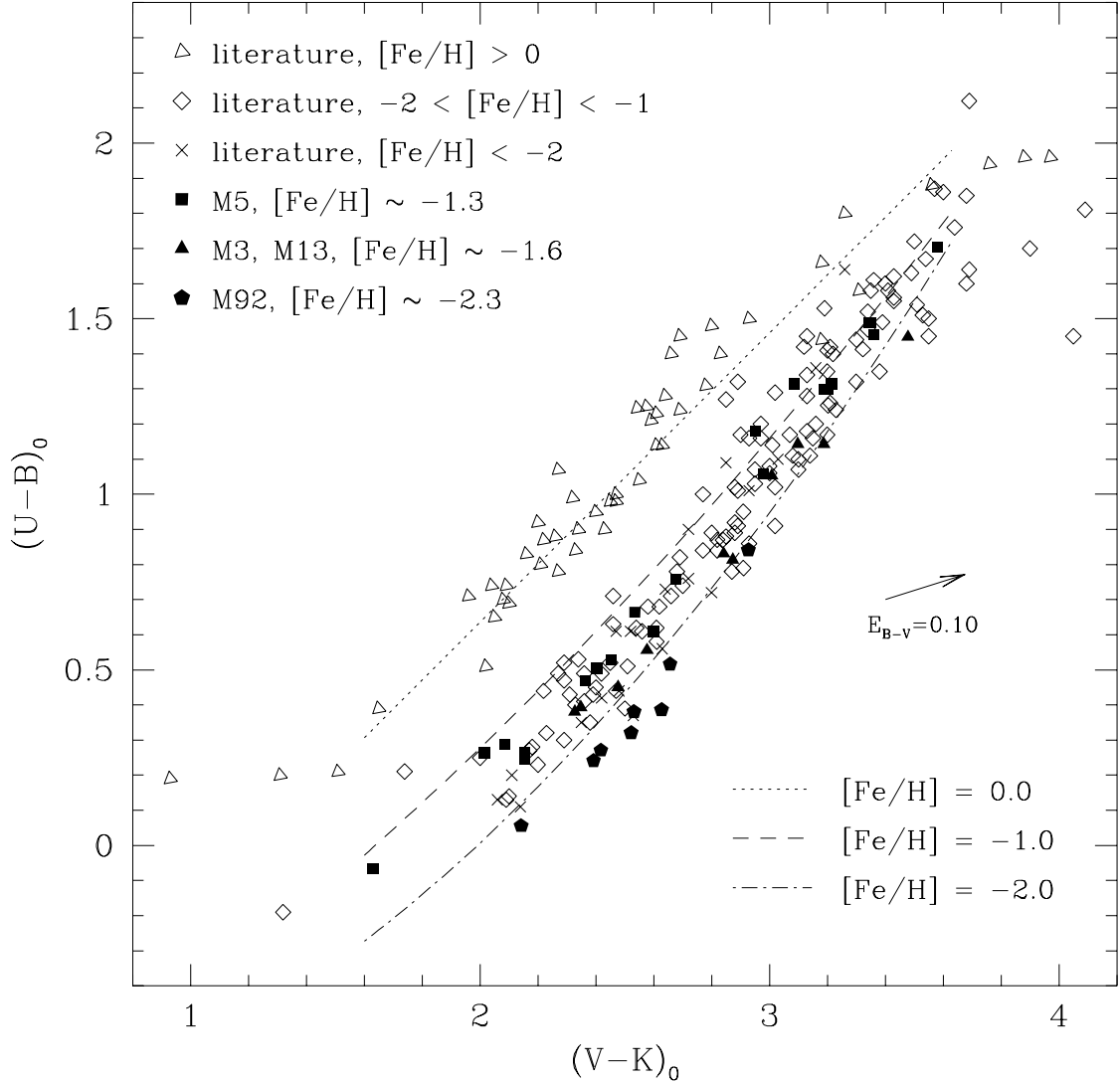}
\caption{The color plane $(U-B)_{0}$ versus $(V-K)_{0}$.
Program stars are
marked with solid symbols as keyed. Literature data are also plotted,
with different symbols for stars with [Fe/H] \ $>0$, [Fe/H] \ $<-2$, and
[Fe/H] between $-1$ and $-2$.  Stars with metallicities between [Fe/H] \ $=0$
and $-1$ have been omitted for clarity.
Fits for [Fe/H] = 0.0, -1.0, -2.0 are superimposed
for the range in which they are valid.
(cf. equation (7)).
The reddening vector indicates the direction along which 
reddening increases in this color-color plot. The length of 
the vector corresponds to $E(B-V)=0.10$.
The strong separation between
giants of differing abundances is clearly visible.
For the program stars, targets
in M92, the metal-poorest cluster in this figure,
lie on a sequence toward bluer
$(U-B)_{0}$. This trend is less obvious but present in the literature
collection. Sequences of all metallicities converge at the coolest
temperatures. \label{ubvk} }
\end{figure}

\begin{figure}
\plotone{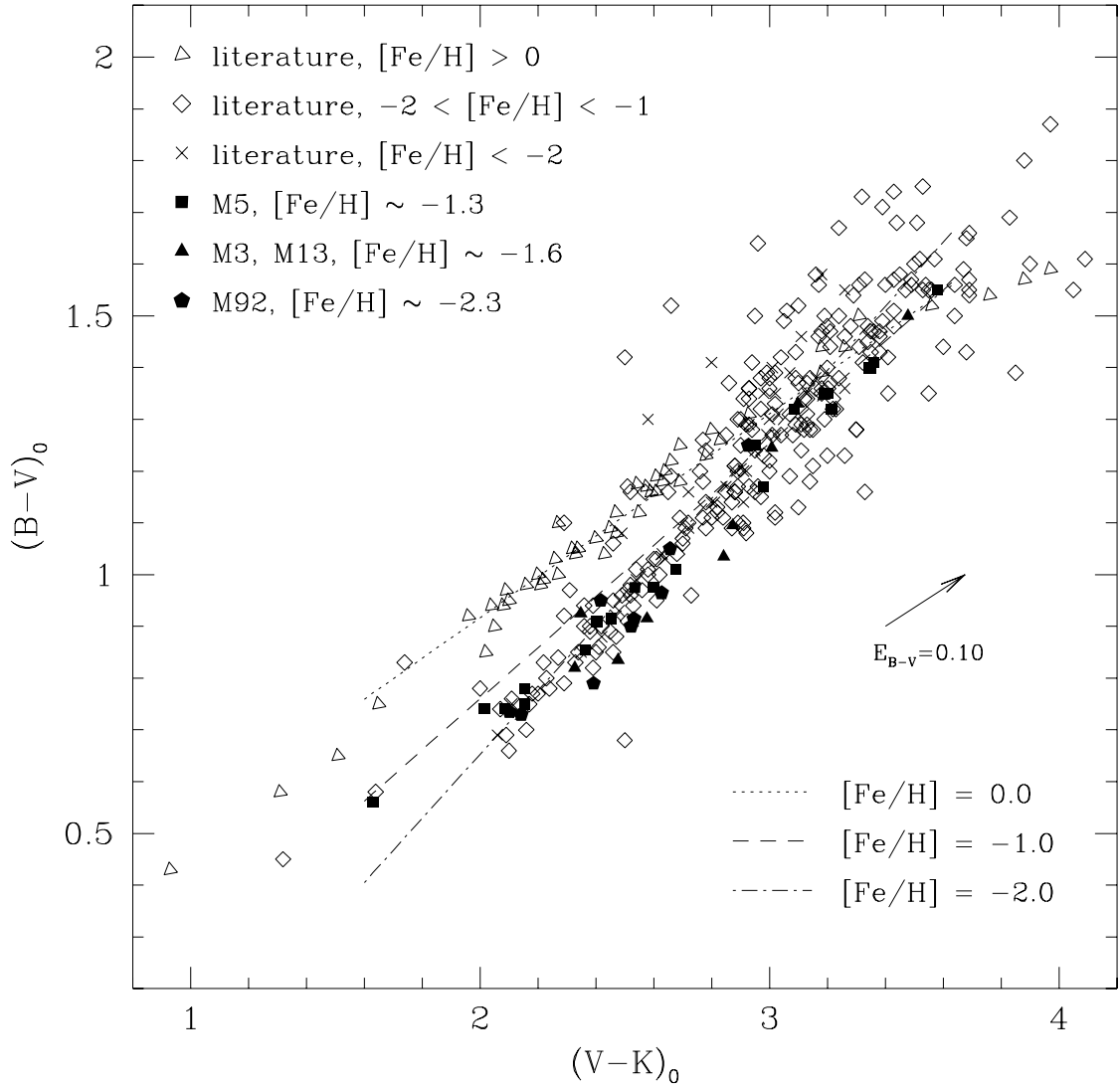}
\caption{The color plane $(B-V)_{0}$
versus $(V-K)_{0}$. Symbols, line types, and reddening vector as in
Fig.~\ref{ubvk}. The metallicity dependence (cf. equation (8)) is clearly
visible. Fits for all [Fe/H] values seem to intersect at
$(V-K)_{0} \sim 3$. \label{bvvk} }
\end{figure}

\begin{figure}
\plotone{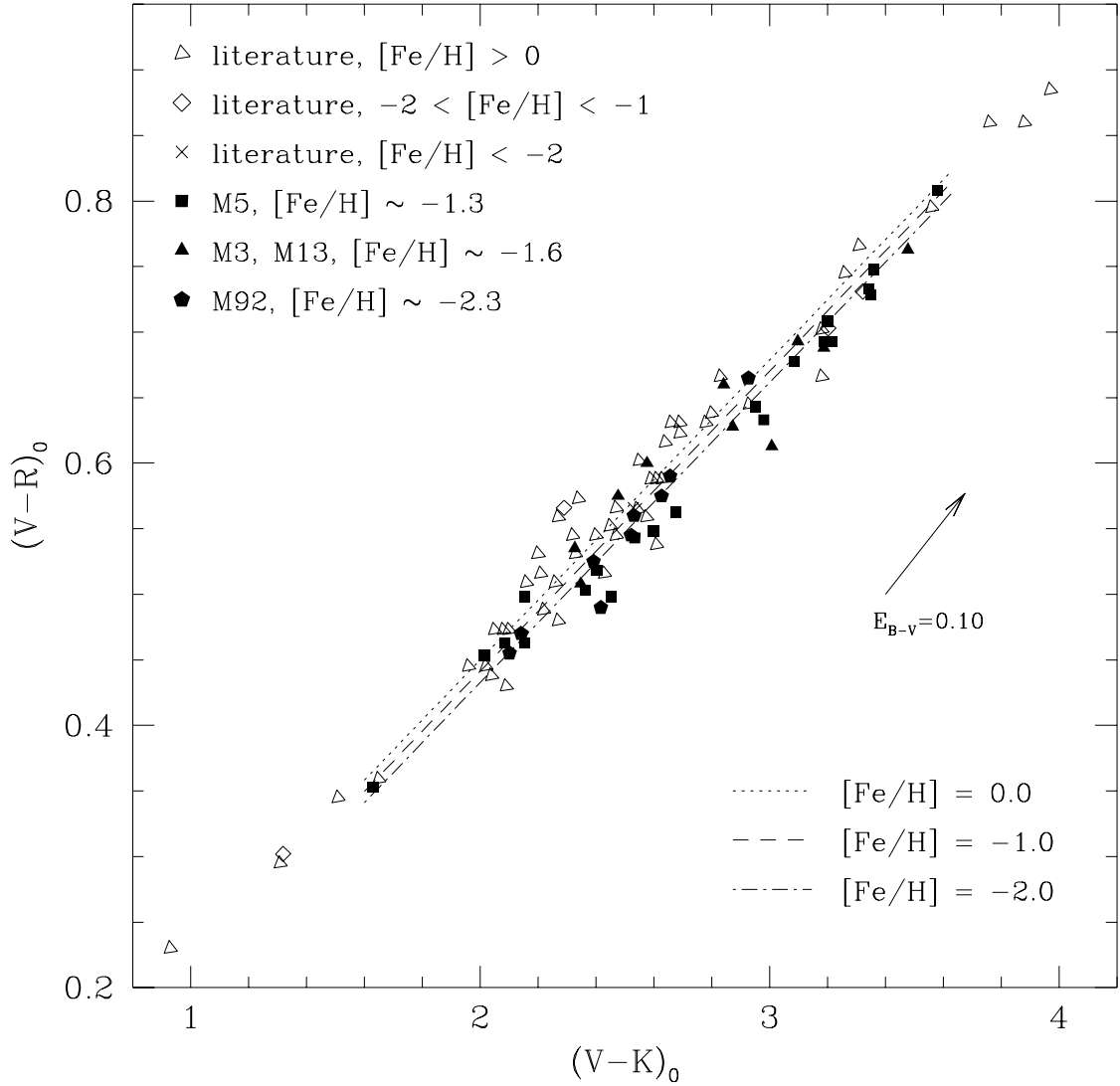}
\caption{The color plane $(V-R)_{0}$
versus $(V-K)_{0}$. Symbols, line types, and reddening vector as in
Fig.~\ref{ubvk}.
There are very few metal-poor stars available from
the literature for this color plane. Only little evidence is visible for a
metallicity trend even though stars with more than a factor of 100
difference in abundance are plotted. Note, however, the subtle
divergence of the $(V-R)_{0}$ sequence for $2 < (V-K)_{0} < 3$. 
For the analytical form of the fits seen in this figure, 
see equation (9). \label{vrvk} }
\end{figure}

\begin{figure}
\plotone{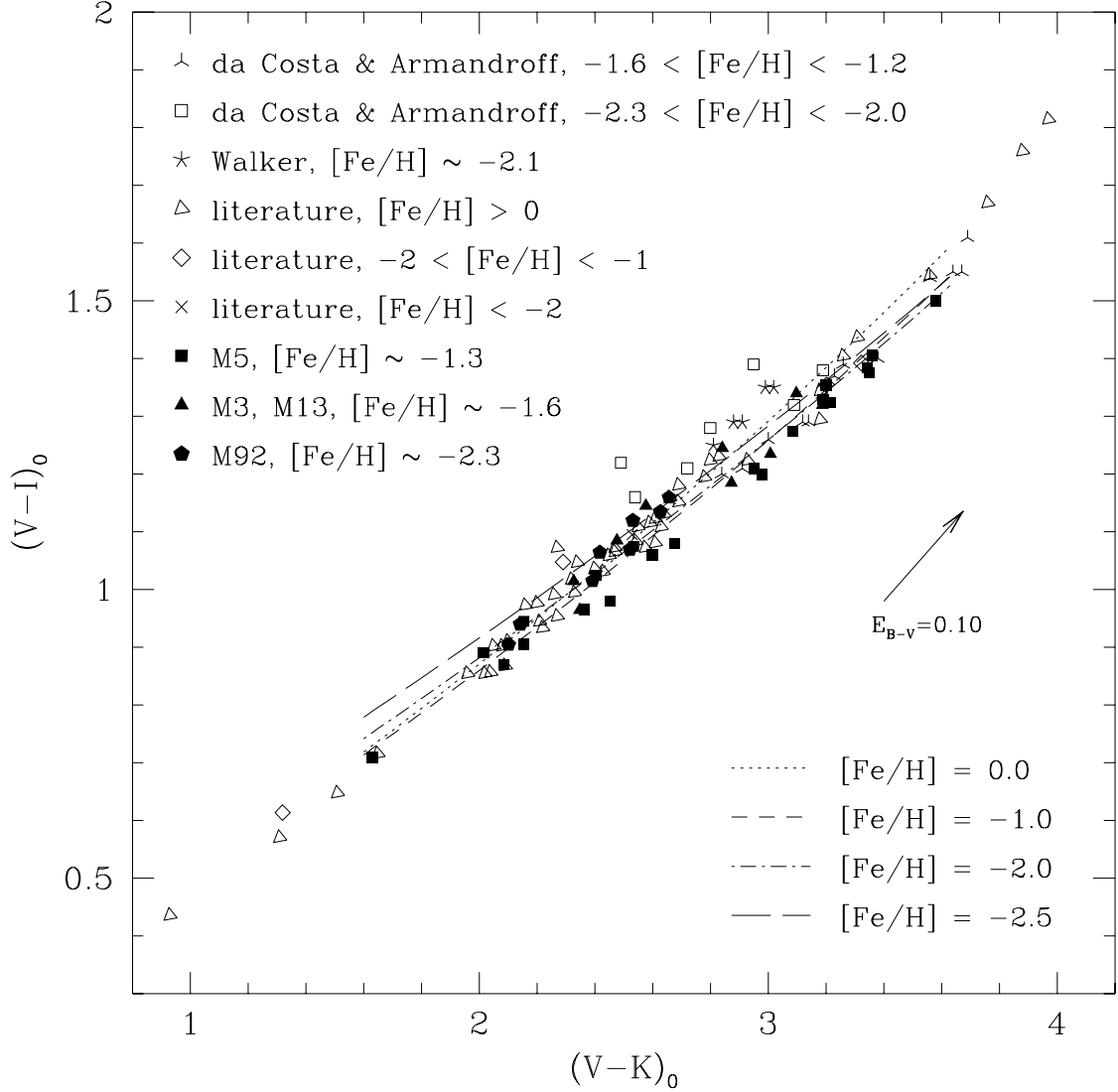}
\caption{The color plane $(V-I)_{0}$
versus $(V-K)_{0}$. Symbols, line types, and reddening vector as in
Fig.~\protect\ref{ubvk}, with three extra symbol types for the main sources of
literature $(V-I)_{0}$ data: Walker (1994) for M68, and da Costa \& Armandroff
(1990) \protect\markcite{walk94,da90} for NGC 6397, M15, NGC 1851, and
NGC 6752.
The three high-lying da Costa \&
Armandroff points all belong to cluster M15 ([Fe/H] \ $=-2.22$)
(Harris, 1996). Note the interesting behavior of $(V-I)_{0}$ with
[Fe/H] (cf. equation (10)). \label{vivk} }
\end{figure}

\begin{figure}
\plotone{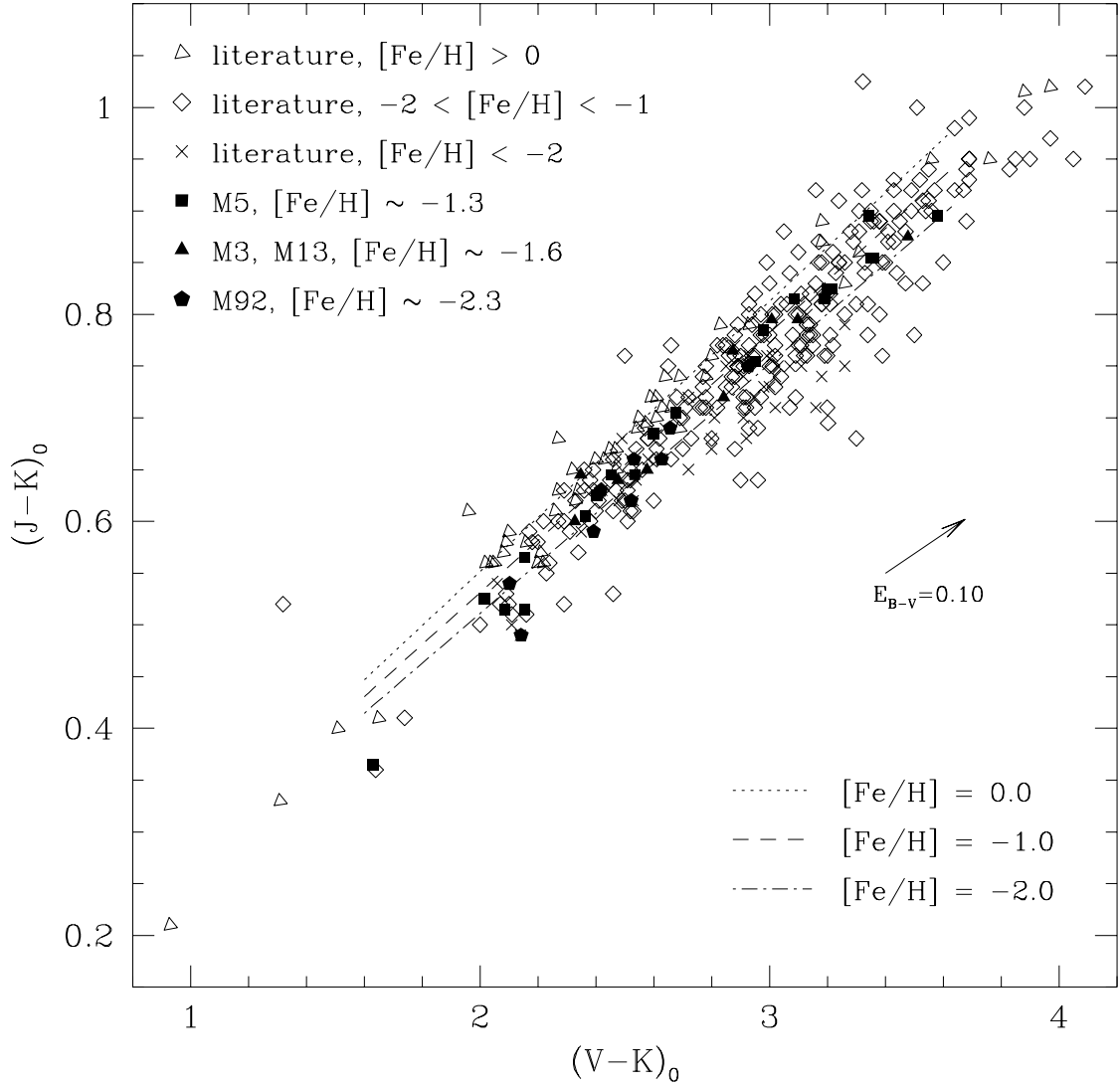}
\caption{The color plane $(J-K)_{0}$
versus $(V-K)_{0}$. Symbols, line types, and reddening vector as in
Fig.~\ref{ubvk}.
We find a statistically significant dependence upon
metallicity (see text and equation (11)).  \label{jkvk} }
\end{figure}

\begin{figure}
\plotone{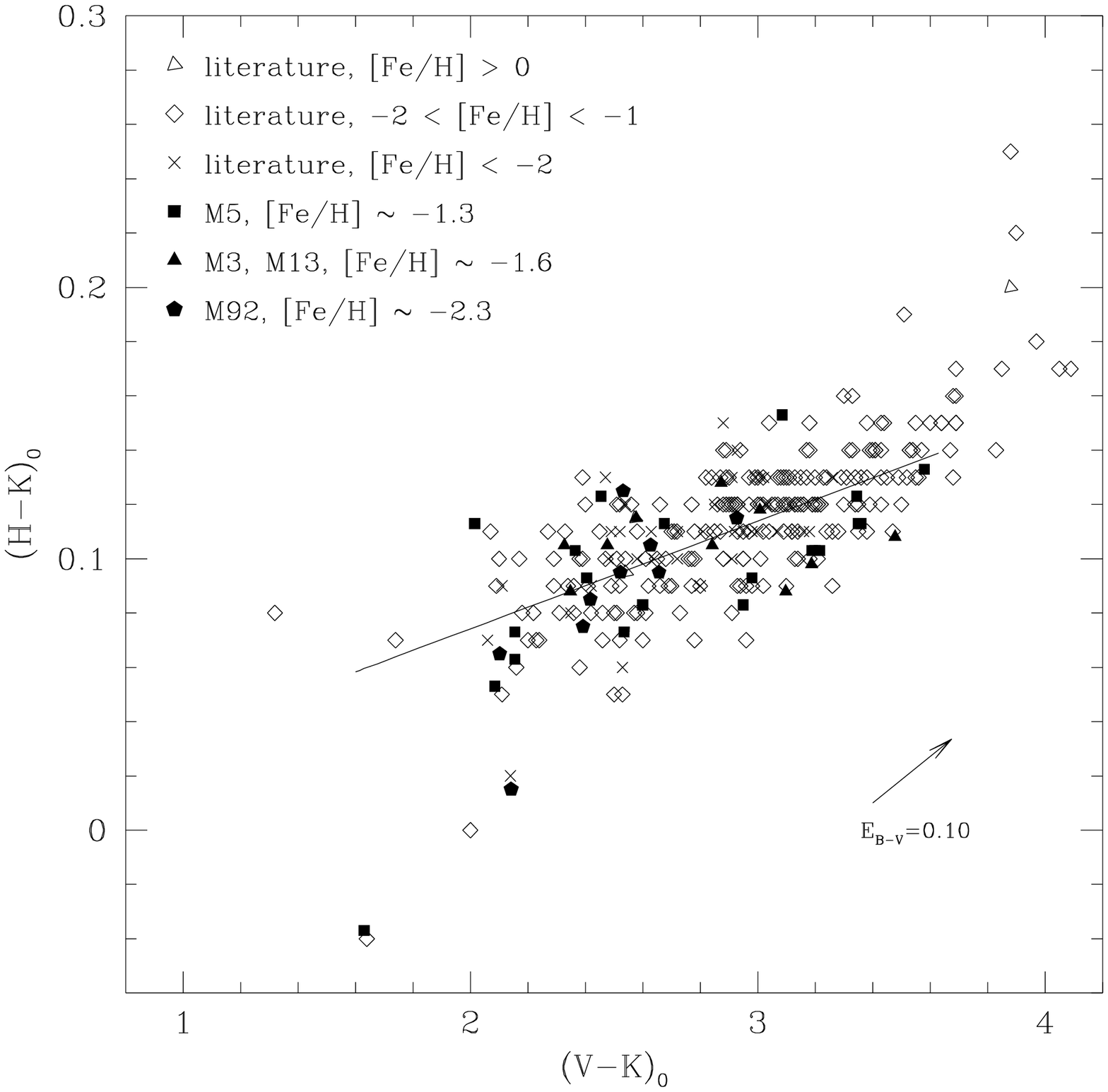}
\caption{The color plane $(H-K)_{0}$
versus $(V-K)_{0}$. Symbols and reddening vector as in Fig.~\ref{ubvk}.
Note that the $(H-K)_{0}$ range is
small compared to its observational error. We supply the
metallicity independent fit for the data (cf. equation (12)).
\label{hkvk} }
\end{figure}

We have presented color sequences and analytical color-color 
fits for giants as a function of [Fe/H]. 
The well-known metallicity sensitivity of $(U-B)_{0}$ is
clearly visible in Fig.~\ref{ubvk}. Our data underscore
this high sensitivity, especially in the higher $T_{\rm eff}$
regime.
A somewhat weaker 
dependence upon [Fe/H] was found for $(B-V)_{0}$. This
color shows an interesting metallicity degeneracy at 
$(B-V)_{0}\sim 1.3$. The colors $(V-R)_{0}$, $(V-I)_{0}$,
and $(J-K)_{0}$ are much less influenced by metallicity, while
$(H-K)_{0}$ is solely a function of $(V-K)_{0}$ (and thus 
$T_{\rm eff}$). The dependence of $(J-K)_{0}$ and $(V-I)_{0}$ 
on [Fe/H] was a surprising result; more data would 
certainly be useful to confirm our findings. 
The extremely weak metallicity dependences of $(V-R)_{0}$ 
and $(V-I)_{0}$ make these viable temperature indicators, 
with the 
advantage of only requiring an optical detector as opposed 
to an optical/IR combination.

\acknowledgments

This research was funded in part by NASA through grant HF-1066.01-94A
from the Space Telescope Science Institute, which is operated by the
Association of Universities for Research in Astronomy, Inc., under
NASA contract NAS5-26555. Thanks to Brent Fisher for a good deal of
hard work in mining the gold out of the literature photometry ore.
We would also like to thank Mario Mateo for his helpful
comments regarding DoPHOT and aperture corrections.
Finally, thanks to the anonymous referee for his/her insightful comments
and suggestions.


 
\newpage
 
 
\begin{deluxetable}{llllll}
\tablewidth{0pt}
\tablecaption{Calibration RMS Errors \label{table1}}
\tablehead{
\colhead{}      & \colhead{V} &
\colhead{B-V} & \colhead{U-B} &
\colhead{V-R} & \colhead{V-I} }
 
\startdata
Night 1 & 0.0165 & 0.0088 & 0.0254 & 0.0047 & 0.0082 \nl
Night 2 & 0.0166 & 0.0072 & 0.0234 & 0.0051 & 0.0052 \nl
Night 3 & 0.0097 & 0.0105 & 0.0187 & 0.0077 & 0.0075 \nl
\enddata
\end{deluxetable}
 
\newpage
 
 
\begin{deluxetable}{lllllllll}
\tablewidth{0pt}
\tablecaption{Photometry Results \label{table2}}
\tablehead{
\colhead{Cluster}      & \colhead{Star ID} &
\colhead{V}      & 
\colhead{U-B} & \colhead{B-V} &
\colhead{V-R} & \colhead{V-I} &
\colhead{Err1}  & \colhead{Err2}}
 
 
\startdata
 
M3\tablenotemark{a}     & II-18    & 14.105 & 0.400  & 0.935  & 0.515  & 0.980 & 0.015 & \nodata   \nl
           & I-21\tablenotemark{c}    & 13.075 & 1.150 & 1.340  & 0.700  & 1.355 & 0.030  & \nodata    \nl
           & III-77    & 13.365 & 1.060  & 1.255  & 0.620  & 1.250 & 0.015 &   0.03u   \nl
           & III-28    & 12.815 & 1.150  & 1.355  & 0.695  & 1.335 & 0.015 &   0.03ui  \nl
           & II-46    & 12.755 & 1.455  & 1.510  & 0.770  & \nodata & 0.015 &   0.03u   \nl
           & IV-25    & 13.680 & 0.820  & 1.105  & 0.630  & 1.200 & 0.015 & \nodata    \nl
M92\tablenotemark{a}    & IV-2\tablenotemark{c}     & 13.500 & 0.285  & 0.970  & 0.505  & 1.095 & 0.030 & \nodata      \nl
           & IV-10    & 13.455 & 0.335  & 0.920  & 0.560  & 1.100 & 0.015 & \nodata     \nl
           & IV-114   & 13.865 & 0.255  & 0.810  & 0.540  & 1.045 & 0.015 & \nodata     \nl
           & III-82    & 13.375 & 0.395  & 0.935  & 0.575  & 1.150 & 0.015 & \nodata     \nl
           & II-70    & 13.120 & 0.400  & 0.985  & 0.590  & 1.165 & 0.015 & \nodata     \nl
           & III-4     & 14.155 & \nodata  & 0.755  & 0.470  & 0.935 & 0.015 & \nodata     \nl
           & XI-19    & 12.870 & 0.530  & 1.070  & 0.605  & 1.190 & 0.015 &   0.03ui  \nl
           & X-49    & 12.220 & 0.855  & 1.270  & 0.680  & \nodata & 0.015 & \nodata    \nl
           & VIII-43    & 14.615 & 0.070  & 0.750  & 0.485  & 0.970 & 0.015 & \nodata    \nl
M13\tablenotemark{a}    & I-24    & 12.955 & 0.845  & 1.055  & 0.675  & 1.275 & 0.015 & \nodata    \nl
           & I-23    & 13.200 & 0.570  & 0.935  & 0.615  & 1.175 & 0.015 & \nodata   \nl
           & I-18    & 13.950 & 0.465  & 0.855  & 0.590  & 1.115 & 0.015 & \nodata   \nl
           & I-2     & 14.290 & 0.395  & 0.840  & 0.550  & 1.045 & 0.015 & \nodata   \nl
M5\tablenotemark{a}     & II-50    & 13.880 & 0.630  & 1.005  & 0.570  & 1.105 & 0.015 &   0.02v   \nl
           & III-3     & 12.470 & 1.475  & 1.440  & 0.770  & 1.450 & 0.015 & \nodata   \nl
           & III-16    & 14.235 & 0.265  & 0.780  & 0.485  & 0.950 & 0.015 & \nodata   \nl
           & III-53    & 13.545 & 0.525  & 0.940  & 0.540  & 1.070 & 0.015 & \nodata   \nl
           & III-56    & 13.365 & 0.685  & 1.005  & 0.565  & 1.120 & 0.015 & \nodata   \nl
           & III-78    & 12.650 & 1.320  & 1.380  & 0.730  & 1.400 & 0.015 & \nodata   \nl
           & IV-19    & 12.610 & 1.320  & 1.380  & 0.715  & 1.375 & 0.015 & \nodata   \nl
           & IV-3   & 14.955 & 0.285  & 0.810  & 0.520  & 0.990 & 0.015 &  \nodata  \nl
           & IV-28    & 14.395 & 0.550  & 0.945  & 0.520  & 1.025 & 0.015 & \nodata   \nl
           & IV-86    & 14.970 & -0.045 & 0.590  & 0.375  & 0.755 & 0.015 &   0.03u   \nl
           & IV-81    & 12.280 & 1.725  & 1.580  & 0.830  & 1.545 & 0.015 & \nodata   \nl
           & IV-47    & 12.425 & 1.510  & 1.430  & 0.755  & 1.430 & 0.015 & \nodata   \nl
           & IV-59    & 12.690 & 1.200  & 1.280  & 0.665  & 1.255 & 0.015 & \nodata   \nl
           & I-25    & 13.595 & 0.780  & 1.040  & 0.585  & 1.125 & 0.015 & \nodata   \nl
           & I-20    & 12.555 & 1.335  & 1.350  & 0.700  & 1.320 & 0.015 & \nodata   \nl
           & I-14    & 13.010 & 1.080  & 1.200  & 0.655  & 1.245 & 0.015 & \nodata   \nl
           & I-68    & 12.500 & 1.510  & 1.430  & 0.750  & 1.420 & 0.015 & \nodata   \nl
           & I-67    & 13.985 & 0.310  & 0.770  & 0.485  & 0.915 & 0.015 & \nodata   \nl
           & I-55    & 13.645 & 0.490  & 0.885  & 0.525  & 1.010 & 0.015 & \nodata   \nl
           & I-1     & 14.185 & 0.285  & 0.770  & 0.475  & 0.935 & 0.015 & \nodata   \nl
           & II-9     & 12.595 & 1.335  & 1.350  & 0.715  & 1.370 & 0.015 &   0.03v   \tablebreak
NGC 6791\tablenotemark{b}  & R17    & 14.575 & 1.770  & 1.530  & 0.800  & 1.510 & 0.010 & \nodata   \nl
           & R4     & 14.000 & 1.415  & 1.580  & 1.355  & \nodata & 0.010 &   0.015r  \nl
           & R19    & 14.140 & 1.975  & 1.630  & 0.895  & 1.740 & 0.010 & \nodata   \nl
           & R23    & 14.875 & 1.520  & 1.385  & 0.700  & 1.305 & 0.010 & \nodata   \nl
           & R9     & 14.130 & 1.985  & 1.595  & 0.885  & 1.705 & 0.010 & \nodata   \nl
           & R24    & 14.980 & 1.750  & 1.450  & 0.785  & 1.460 & 0.010 & \nodata   \nl
           & R22    & 14.525 & 1.580  & 1.390  & 0.710  & 1.355 & 0.015 & \nodata   \nl
           & R12    & 13.835 & 2.040  & 1.650  & 0.930  & 1.920 & 0.010 &   0.015r  \nl
           & R11    & 14.600 & 1.765  & 1.500  & 0.775  & 1.475 & 0.010 &   0.02b   \nl
           & R3     & 14.070 & 2.220  & 1.710  & 0.965  & 1.940 & 0.015 &   0.03u   \nl
           & R21    & 14.720 & 1.700  & 1.470  & 0.770  & 1.445 & 0.015 &   0.03u   \nl
           & R25    & 14.715 & 1.490  & 1.395  & 0.715  & 1.340 & 0.015 & \nodata  \nl
           & R10    & 14.570 & 1.765  & 1.585  & 0.800  & 1.540 & 0.015 & \nodata  \nl
           & R16    & 13.740 & 2.055  & 1.635  & 0.865  & 1.730 & 0.015 & \nodata

\enddata
 
\tablenotetext{a}{The star IDs follow the numbering scheme used by Cohen et al. (1978)\markcite{cohen78}.}
\tablenotetext{b}{The star IDs follow the numbering scheme used by Garnavich et al. (1994)\markcite{gvzh94}.}
\tablenotetext{c}{These stars were only observed during night 1.}
 
\tablecomments{All values in this table are rounded to the nearest 0.005.}
 
\end{deluxetable}

\newpage
 
 
\begin{deluxetable}{lllllllllllll}
\tablewidth{0pt}
\tablenum{3}
\tablecaption{Photometry Results - Dereddened \label{table3}}
\tablehead{
\colhead{Cluster}      & \colhead{Star ID} &
\colhead{[Fe/H]\tablenotemark{1,2}} & 
\colhead{$E_{B-V}$\tablenotemark{1,3}} & \colhead{V}      & 
\colhead{U-B} & \colhead{B-V} &
\colhead{V-R} & \colhead{V-I} &
\colhead{V-K} & \colhead{K} &
\colhead{J-K} & \colhead{H-K} }

\startdata

                                
 M3 &  II-18 & -1.57 & 0.01 & 14.075 &  0.393 &   0.925 &   0.508 &   0.965 &   2.348 &   11.727 &   0.645 &   0.088 \nl    
    &  I-21 &        &      & 13.045 &  1.143 &   1.330 &   0.693 &   1.340 &   3.098 &    9.947 &   0.795 &   0.088 \nl    
    &  III-77 &      &      & 13.335 &  1.053 &   1.245 &   0.613 &   1.235 &   3.008 &   10.327 &   0.795 &   0.118 \nl    
    &  III-28 &      &      & 12.785 &  1.143 &   1.345 &   0.688 &   1.320 &   3.188 &    9.597 &   0.815 &   0.098 \nl    
    &  II-46 &       &      & 12.725 &  1.448 &   1.500 &   0.763 & \nodata &   3.478 &    9.247 &   0.875 &   0.108 \nl    
    &  IV-25 &       &      & 13.650 &  0.813 &   1.095 &   0.628 &   1.185 &   2.873 &   10.777 &   0.765 &   0.128 \nl    
 M92 & IV-2 & -2.29 & 0.02 & 13.440 &  0.271 &   0.950 &   0.490 &   1.065 &   2.417 &   11.023 &   0.630 &   0.085 \nl    
     & IV-10 &       &      & 13.395 &  0.321 &   0.900 &   0.545 &   1.070 &   2.522 &   10.873 &   0.620 &   0.095 \nl    
     & IV-114 &      &      & 13.805 &  0.241 &   0.790 &   0.525 &   1.015 &   2.392 &   11.413 &   0.590 &   0.075 \nl    
     & III-82 &      &      & 13.315 &  0.381 &   0.915 &   0.560 &   1.120 &   2.532 &   10.783 &   0.660 &   0.125 \nl    
     & II-70 &       &      & 13.060 &  0.386 &   0.965 &   0.575 &   1.135 &   2.627 &   10.433 &   0.660 &   0.105 \nl    
     & III-4 &       &    & 14.095 &  \nodata &   0.735 &   0.455 &   0.905 &   2.102 &   11.993 &   0.540 &   0.065 \nl    
     & XI-19 &       &      & 12.810 &  0.516 &   1.050 &   0.590 &   1.160 &   2.657 &   10.153 &   0.690 &   0.095 \nl    
     & X-49 &        &      & 12.160 &  0.841 &   1.250 &   0.665 & \nodata &   2.927 &    9.233 &   0.750 &   0.115 \nl    
     & VIII-43 &     &      & 14.555 &  0.056 &   0.730 &   0.470 &   0.940 &   2.142 &   12.413 &   0.490 &   0.015 \nl    
 M13 & I-24 & -1.54 & 0.02 & 12.895 &  0.831 &   1.035 &   0.660 &   1.245 &   2.842 &   10.053 &   0.720 &   0.105 \nl    
     & I-23 &        &     & 13.140 &  0.556 &   0.915 &   0.600 &   1.145 &   2.577 &   10.563 &   0.650 &   0.115 \nl    
     & I-18 &        &     & 13.890 &  0.451 &   0.835 &   0.575 &   1.085 &   2.477 &   11.413 &   0.640 &   0.105 \nl    
     & I-2 &         &     & 14.230 &  0.381 &   0.820 &   0.535 &   1.015 &   2.327 &   11.903 &   0.600 &   0.105 \nl    
 M5 &  I-1 & -1.29 & 0.03 & 14.095 &  0.264 &   0.740 &   0.453 &   0.890 &   2.015 &   12.080 &   0.525 &   0.113 \nl    
    &  I-14 &         &    & 12.920 &  1.059 &   1.170 &   0.633 &   1.200 &   2.980 &    9.940 &   0.785 &   0.093 \nl    
    &  I-20 &         &    & 12.465 &  1.314 &   1.320 &   0.678 &   1.275 &   3.085 &    9.380 &   0.815 &   0.153 \nl    
    &  I-25 &         &    & 13.505 &  0.759 &   1.010 &   0.563 &   1.080 &   2.675 &   10.830 &   0.705 &   0.113 \nl    
    &  I-55 &         &    & 13.555 &  0.469 &   0.855 &   0.503 &   0.965 &   2.365 &   11.190 &   0.605 &   0.103 \nl    
    &  I-67 &         &    & 13.895 &  0.289 &   0.740 &   0.463 &   0.870 &   2.085 &   11.810 &   0.515 &   0.053 \nl    
    &  I-68 &         &    & 12.410 &  1.489 &   1.400 &   0.728 &   1.375 &   3.350 &    9.060 &   0.855 &   0.113 \nl    
    &  II-9 &         &    & 12.505 &  1.314 &   1.320 &   0.693 &   1.325 &   3.215 &    9.290 &   0.825 &   0.103 \nl    
    &  II-50 &        &    & 13.790 &  0.609 &   0.975 &   0.548 &   1.060 &   2.600 &   11.190 &   0.685 &   0.083 \nl    
    &  III-3 &        &    & 12.380 &  1.454 &   1.410 &   0.748 &   1.405 &   3.360 &    9.020 &   0.855 &   0.113 \nl    
    &  III-16 &       &    & 14.145 &  0.244 &   0.750 &   0.463 &   0.905 &   2.155 &   11.990 &   0.515 &   0.063 \nl    
    &  III-53 &       &    & 13.455 &  0.504 &   0.910 &   0.518 &   1.025 &   2.405 &   11.050 &   0.625 &   0.093 \nl    
    &  III-56 &       &    & 13.275 &  0.664 &   0.975 &   0.543 &   1.075 &   2.535 &   10.740 &   0.645 &   0.073 \nl    
    &  III-78 &       &    & 12.560 &  1.299 &   1.350 &   0.708 &   1.355 &   3.200 &    9.360 &   0.825 &   0.103 \nl    
    &  IV-3 &         &    & 14.865 &  0.264 &   0.780 &   0.498 &   0.945 &   2.155 &   12.710 &   0.565 &   0.073 \nl    
    &  IV-19 &        &    & 12.520 &  1.299 &   1.350 &   0.693 &   1.330 &   3.190 &    9.330 &   0.815 &   0.103 \nl    
    &  IV-28 &        &    & 14.305 &  0.529 &   0.915 &   0.498 &   0.980 &   2.455 &   11.850 &   0.645 &   0.123 \nl    
    &  IV-47 &        &    & 12.335 &  1.489 &   1.400 &   0.733 &   1.385 &   3.345 &    8.990 &   0.895 &   0.123 \nl    
    &  IV-59 &        &    & 12.600 &  1.179 &   1.250 &   0.643 &   1.210 &   2.950 &    9.650 &   0.755 &   0.083 \nl    
    &  IV-81 &        &    & 12.190 &  1.704 &   1.550 &   0.808 &   1.500 &   3.580 &    8.610 &   0.895 &   0.133 \nl    
    &  IV-86 &        &    & 14.880 & -0.066 &   0.560 &   0.353 &   0.710 &   1.630 &   13.250 &   0.365 &  -0.037 \tablebreak
 NGC 6791 & R17   & 0.20 & 0.13 & 14.175 &  1.677 &  1.395 &   0.700 &   1.309 & \nodata &  \nodata & \nodata & \nodata \nl    
       & R4 &         &     & 13.600 &  1.322 &   1.445 &   1.255 & \nodata & \nodata &  \nodata & \nodata & \nodata \nl    
       & R19 &        &     & 13.740 &  1.882 &   1.495 &   0.795 &   1.539 & \nodata &  \nodata & \nodata & \nodata \nl    
       & R23 &        &     & 14.475 &  1.427 &   1.250 &   0.600 &   1.104 & \nodata &  \nodata & \nodata & \nodata \nl    
       & R9 &         &     & 13.730 &  1.892 &   1.460 &   0.785 &   1.504 & \nodata &  \nodata & \nodata & \nodata \nl    
       & R24 &        &     & 14.580 &  1.657 &   1.315 &   0.685 &   1.259 & \nodata &  \nodata & \nodata & \nodata \nl    
       & R22 &        &     & 14.125 &  1.487 &   1.255 &   0.610 &   1.154 & \nodata &  \nodata & \nodata & \nodata \nl    
       & R12 &        &     & 13.435 &  1.947 &   1.515 &   0.830 &   1.719 & \nodata &  \nodata & \nodata & \nodata \nl    
       & R11 &        &     & 14.200 &  1.672 &   1.365 &   0.675 &   1.274 & \nodata &  \nodata & \nodata & \nodata \nl    
       & R3 &         &     & 13.670 &  2.127 &   1.575 &   0.865 &   1.739 & \nodata &  \nodata & \nodata & \nodata \nl    
       & R21 &        &     & 14.320 &  1.607 &   1.335 &   0.670 &   1.244 & \nodata &  \nodata & \nodata & \nodata \nl    
       & R25 &        &     & 14.315 &  1.397 &   1.260 &   0.615 &   1.139 & \nodata &  \nodata & \nodata & \nodata \nl    
       & R10 &        &     & 14.170 &  1.672 &   1.450 &   0.700 &   1.339 & \nodata &  \nodata & \nodata & \nodata \nl    
       & R16 &        &     & 13.340 &  1.962 &   1.500 &   0.765 &   1.529 & \nodata &  \nodata & \nodata & \nodata

\enddata
 
\tablenotetext{1}{All globular cluster data: Harris (1996). \markcite{harris96}}
 
\tablenotetext{2}{For NGC 6791: Friel and Janes (1993)\markcite{fj93} and 
Garnavich et al. (1994).\markcite{gvzh94} }
 
\tablenotetext{3}{For NGC 6791: Friel and Janes (1993),\markcite{fj93} 
Harris and Canterna (1981),\markcite{hc81}
and Liebert et al. (1994).\markcite{liebert94}
We would like to note, however, that higher 
values for $E_{B-V}$ were obtained by Demarque 
et al. (1992)\markcite{demarque92} and Kaluzny (1990).\markcite{kal90} }

\tablecomments{The reddening calculations in this table are based on 
Cardelli et al. (1989).}
\tablecomments{The star numbering system used is the same as the one
in Table 2.}
\tablecomments{The dereddened infrared magnitudes and colors were obtained
by applying the Cardelli et al. reddening curve to observed photometry by
Frogel et al. (1983)\markcite{frogel83}
and Cohen et al. (1978).\markcite{cohen78}
For M3, M13, and M92, Cohen et al. cite infrared photometry 
errors of 0.02 mag for $K$, $J-H$, and $H-K$.  
The infrared photometry errors for M5 quoted by Frogel et al.
are 0.03 mag for $K$, $J-K$, and $H-K$.}
 
\end{deluxetable}


\newpage

\newpage

\newpage

\newpage

\newpage

\newpage

\newpage

\begin{references}

\reference{arp55} Arp, H. C. 1955, \aj, 60, 317
\reference{arp62} Arp, H. C. 1962, \apj, 135, 311
\reference{bg78} Bell, R. A., \& Gustafsson, B. 1978, \aaps, 34, 229
\reference{bg89} Bell, R. A., \& Gustafsson, B. 1989, \mnras, 236, 653
\reference{b79} Bessell, M. S. 1979, \pasp, 91, 589
\reference{bbsw89} Bessell, M. S., Brett, J. M., Scholz, M., \& Wood,
P. R. 1989, \aaps, 77, 1
\reference{bbsw89} Bessell, M. S., Brett, J. M., Scholz, M., \& Wood,
P. R. 1991, \aaps, 89, 335
\reference{bk79} Buser, R. \& Kurucz, R. L. 1979, \aap, 70, 555
\reference{card89} Cardelli, J. A., Clayton, G. C., Mathis, J. S. 1989, 
\apj, 345, 245
\reference{cathey74} Cathey, L. 1974, \aj, 79, 1370
\reference{cds92} Cayrel de Strobel, G., Hauck, B., Fran\c{c}ois, 
P., Th\'{e}venin, F., Friel, E., Mermilliod, M., and Borde, S.
1992, \aaps, 95, 273
\reference{cohen78} Cohen, J. C., Frogel, J. A., and Persson, S. E.
1978, \apj, 222, 165
\reference{c80a} Cousins, A. W. 1980a, South African
Astron. Obs. Circ., 1, 166
\reference{c80b} Cousins, A. W. 1980b, South African
Astron. Obs. Circ., 1, 234
\reference{c96} Corbally, C. 1996, \baas, 189, 78.12
\reference{da90} da Costa, G. S., \& Armandroff, T. E. 1990, \aj, 100, 162
\reference{demarque92} Demarque, P., Green, E. M., and Guenther, 
D. B. 1992, \aj, 103, 151
\reference{d93}  DiBenedetto, G. P. 1993, \aap, 270, 315
\reference{dbvr96} Dyck, H. M., Benson, J. A., van Belle, G. T., and
Ridgway, S. T. 1996, \aj, 111, 1705
\reference{efmn82} Elias, J. H., Frogel, J. A., Matthews, K., and
Neugebauer, G. 1982, AJ, 87, 1029
\reference{fj93} Friel, E. D., and Janes, K. A. 1993, \aap, 267, 75
\reference{frogel83} Frogel, J. A., Persson, S. E., and 
Cohen, J. G. 1983, \apjs, 53, 713
\reference{gvzh94} Garnavich, P. M., VandenBerg, D. A., Zurek, D. R.,
and Hesser, J. E. 1994, \aj, 107, 1097
\reference{g88} Green, E. M. 1988, in Calibration of Stellar Ages,
ed. A. G. Davis Philip (Schenectady, NY: L. Davis), 81
\reference{RYI} Green, E. M., Demarque, P., \& King, C. R. 1987, The
Revised Yale Isochrones and Luminosity Functions (New Haven: Yale
University Observatory)
\reference{hc81} Harris, W. E., \& Canterna, R. 1981, \aj, 86, 1332
\reference{harris96} Harris, W. E. 1996, \aj, 112, 1487
\reference{kal90} Kaluzny, J. 1990, \mnras, 243, 492
\reference{kal95} Kaluzny, J., Rucinski, S. M. 1995, \aaps, 114, 1
\reference{kinman65} Kinman, T. D. 1965, \apj, 142, 655
\reference{k92} Kurucz, R. L. 1992, in IAU Symp. 149, The Stellar
Populations of Galaxies, ed. B. Barbuy \& A. Renzini (Dordrecht:
Kluwer), 225
\reference{j66} Johnson, H. L. 1966, \araa, 4, 193
\reference{land92} Landolt, A. 1992, \aj, 104, 340
\reference{liebert94} Liebert, J., Saffer, R. A., and Green, 
E. M. 1994, \aj, 107, 1408
\reference{mcwilliam90} McWilliam, A. 1990, \apjs, 74, 1075
\reference{mm78} Morel, M., \& Magnenat, P. 1978, \aaps, 34, 477
\reference{m92} Mould, J. R. 1992, in IAU Symp. 149, The Stellar
Populations of Galaxies, ed. B. Barbuy \& A. Renzini (Dordrecht:
Kluwer), 181
\reference{press92} Press, W., Teukolsky, S., Vetterling, W., 
Flannery, B. 1992, Numerical Recipes, 2nd Edition (Cambridge
University Press)
\reference{rjww80} Ridgway, S. T., Joyce, R. R., White, N. M., and
Wing, R. F. 1980, \apj, 235, 126
\reference{schech93} Schechter, P. L., Mateo, M., and Saha, A. 1993, 
\pasp, 105, 1342.
\reference{tft95} Tiede, G. P., Frogel, J. A., \& Terndrup,
D. M. 1995, \baas, 186, 50.06
\reference{vdb62} van den Bergh, S. 1962, \aj, 67, 486
\reference{walk94} Walker, A. R., 1994, AJ, 108, 555
\reference{wf96} Worthey, G. \& Fisher, B. 1996, \baas, 28, 1366, \#72.05
\reference{w94} Worthey, G. 1994, \apjs, 95, 107

\end{references}
\end{document}